\documentclass{article}

\usepackage{arxiv}

\usepackage[utf8]{inputenc} % allow utf-8 input
\usepackage[T1]{fontenc}    % use 8-bit T1 fonts
\usepackage{hyperref}       % hyperlinks
\usepackage{url}            % simple URL typesetting
\usepackage{booktabs}       % professional-quality tables
\usepackage{amsfonts}       % blackboard math symbols
\usepackage{nicefrac}       % compact symbols for 1/2, etc.
\usepackage{microtype}      % microtypography
\usepackage{lipsum}		% Can be removed after putting your text content
\usepackage{graphicx}
\usepackage{natbib}
\usepackage{doi}
\usepackage{multirow}
\usepackage{amsthm}
\usepackage{array}
\newtheorem{definition}{Definition}

\usepackage{balance} % for balancing columns on the final page
\usepackage{caption}
\usepackage{subcaption}
\usepackage{mathtools}
\usepackage{algorithm}
\usepackage{algpseudocode}

\title{Modelling Cournot Games as Multi-agent Multi-armed Bandits}

\author{ Kshitija Taywade \\
	Department of Computer Science\\
	University of Kentucky\\
	\And
    Brent Harrison \\
	Department of Computer Science\\
	University of Kentucky\\
	
	\And
    Adib Bagh \\
	Department of Economics\\
	University of Kentucky\\
}

\hypersetup{
pdftitle={Modelling Cournot Games as Multi-agent Multi-armed Bandits},
pdfsubject={Multi-armed Bandits},
pdfauthor={Kshitija Taywade, Brent Harrison, Adib Bagh},
pdfkeywords={Multi-agent Learning, Multi-armed Bandits, Cournot Games},
}

\begin{document}
\maketitle

\begin{abstract}

We investigate the use of a multi-agent multi-armed bandit (MA-MAB) setting for modeling repeated Cournot oligopoly games, where the firms acting as agents choose from the set of arms representing production quantity (a discrete value). Agents interact with separate and independent bandit problems. In this formulation, each agent makes sequential choices among arms to maximize its own reward. Agents do not have any information about the environment; they can only see their own rewards after taking an action. However, the market demand is a stationary function of total industry output, and random entry or exit from the market is not allowed. Given these assumptions, we found that an $\epsilon$-greedy approach offers a more viable learning mechanism than other traditional MAB approaches, as it does not require any additional knowledge of the system to operate. We also propose two novel approaches that take advantage of the ordered action space: $\epsilon$-greedy+HL and $\epsilon$-greedy+EL. These new approaches help firms to focus on more profitable actions by eliminating less profitable choices and hence are designed to optimize the exploration. We use computer simulations to study the emergence of various equilibria in the outcomes and do the empirical analysis of joint cumulative regrets.
\end{abstract}


\section{Introduction}

The Cournot oligopoly model \cite{cournot1838recherches} is a well-known model in economic game theory. In this model, firms compete over production quantities, and the total quantity produced by the firms affects the market price. Cournot games are used to model many real-world problems such as energy systems \cite{kirschen2018fundamentals}, transportation networks \cite{bimpikis2019cournot} and healthcare systems \cite{chletsos2019hospitals}. There are three main types of equilibria associated with this model: Walrasian equilibrium, Cournot/Nash equilibrium, and collusive equilibrium. Cournot games have been modeled in several ways in the literature with various assumptions on the firms' cognitive capabilities and rationalities. Different learning mechanisms have been analyzed in order to understand the conditions required to arrive at a specific equilibrium.

Many learning mechanisms from the literature assume that the firms have some knowledge of the market and their competitors. In this paper, we model repeated Cournot games using MA-MAB setting, where firms learn independently and are autonomous. This setting provides a practical framework to model the Cournot game when there are assumptions of low cognitivity for the firms involved. Here, information on competitors is not available to the firms, and they cannot deduce the demand function. Moreover, firms do not even need to know their own cost function necessarily. However, we assume that the random entry and exit of the firms are not allowed. MAB framework facilitates the use of exploration-exploitation approaches. This MA-MAB setting is different from other MA-MAB settings studied in the literature.

Balancing the exploration-exploitation trade-off is an important part of learning in MAB settings. In our MA-MAB setting, as multiple agents learn simultaneously, where learning involves exploration, high uncertainty in the rewards can be seen from the perspective of a single agent having no knowledge of the environment. This uncertainty can vary based on agents' exploration-exploitation strategies. However, it is guaranteed that the rewards are generated using the same mechanism every time, due to the assumption of static demand and the prevention of random entry or exit of firms from the market. We also assume that agents learn simultaneously using the same learning mechanism, and as they learn, their strategies converge. Because of this, we consider these bandits to be stochastic rather than non-stationary or non-stochastic.

In \cite{lattimore2020bandit}, authors suggest that in real-world scenarios, it is unlikely to have truly stochastic bandit models; therefore only their usefulness matters. They encourage examining algorithms' tolerance for deviations from the assumptions that the means of arms stay the same. We investigate three well-known algorithms: $\epsilon$-greedy, UCB, and Thompson sampling. Literature mainly focuses on Bernoulli bandits, while we are dealing with normal bandits with unknown parameters. UCB-Normal \cite{cowan2017normal} which is designed for such problems, fails to converge in our settings. Similarly, Thompson sampling proposed in \cite{honda2014optimality} depends on correctly setting priors to succeed. For both UCB and Thompson sampling, we found that the parameters controlling exploration need to be set cautiously to get any convergence in our setting; however, information about market is needed to set such parameters. Therefore, we think $\epsilon$-greedy remains the most viable option.

In this work, we propose two novel approaches built on top of the $\epsilon$-greedy algorithm that take advantage of the ordered action space to optimize exploration by reducing the action space. The first approach is similar to hierarchical learning; we call it $\epsilon$-greedy+HL. It is based on partitioning the action space into few buckets and choosing the best bucket. The second method resorts to action/arm elimination to reduce the action space, and hence called as $\epsilon$-greedy+EL; EL stands for elimination. In real-world scenarios, the exploration can be costly, and we think that the operating mechanism of our proposed approaches can significantly reduce the exploration cost. Therefore, these newly proposed approaches are more adaptable for real-world usage and applicable to other similar problems with ordered action spaces such as dynamic pricing.

We evaluate these approaches empirically by running different kinds of simulations. We study these approaches in terms of resulting equilibrium, scalability, and sensitivity to hyper-parameters. We investigate Cournot models with symmetric firms (same marginal cost) and extend our study to asymmetric firms (different marginal cost). We consider both deterministic and stochastic inverse demand functions. The stochastic demand function represents small fluctuations in the market demand as it is common in real-world scenarios. In experiments, we start with simulations same as that used in \cite{waltman2008q,xu2020reinforcement}; these are the small-scale simulations as they contain total firms between $2-6$. We further scale the models in terms of the number of firms and the size of the action space. In these evaluations, we compare joint quantity, joint profit, and joint cumulative regret obtained by $\epsilon$-greedy and our proposed methods. In small-scale settings, with a limited number of firms and action choices, collusion in various degrees can be seen. In general, outcomes obtained by our proposed methods are more collusive than baseline $\epsilon$-greedy, although full collusion usually does not emerge. Although the real-world scenarios are more complex than our simulations, the results still indicate the possibility of the emergence of collusive behavior, even without explicit instructions to collude. Joint cumulative regret obtained by $\epsilon$-greedy+HL is the lowest, followed by that obtained by $\epsilon$-greedy+EL. Moreover, our proposed approaches converge much faster than baseline $\epsilon$-greedy method.

\subsection{Related Work}
Cournot oligopoly games, and their convergence to various equilibria, have been widely studied in past research. The literature establishes that the long-term outcome of a Cournot oligopoly model depends on the underlying learning mechanism, firms' rationality, and the memory size of the firms. The learning of economic agents can be widely divided into two categories: individual learning and social learning \cite{vriend2000illustration}. In individual learning models, an agent learns exclusively from its own experience, whereas in social learning models, an agent also learns from the experience of other agents. Learning frameworks also reflect the rationality of agents. Several works have studied individual learning mechanisms \cite{vriend2000illustration,arifovic2006revisiting,vallee2009convergence,fudenberg1998theory,riechmann2006cournot}. Some works claim that while social learning leads to Walrasian equilibrium, individual learning scheme results in convergence of the Nash equilibrium \cite{vriend2000illustration,vega1997evolution,franke1998coevolution,dawid2011adaptive,bischi2015evolutionary}. In this work, we use MA-MAB setting \cite{thompson1933likelihood} as a learning framework. Firms make decisions based on the profit/reward they are getting, mostly unaware of the impact of other firms' actions on the market price; this can be called as implicit individual learning.

There is also vast literature on multi-agent MAB problems. Some papers include models inspired by the cognitive radio problem where agents pulling the same arm cause collision resulting in zero or small rewards \cite{anandkumar2011distributed,liu2010distributed,rosenski2016multi,besson2018multi,hanawal2021multi,wang2020optimal}. Other works study different kinds of MA-MAB models with various degrees of collaboration and communication among agents \cite{sankararaman2019social,chawla2020gossiping,vial2021robust,agarwal2021multi}. To our knowledge, MA-MAB setting has not been studied yet for Cournot games; however, some RL approaches have been applied.

Closely related to our work is the work by \cite{waltman2008q} and \cite{xu2020reinforcement} who have studied RL methods. \citeauthor{waltman2008q} analyze the results of Q-learning in a Cournot game with discrete action space and explain the emergence of collusive behavior. They focus on three types of firms; firms in our experiments are similar to the firms without memory as considered in their paper. \cite{kimbrough2003note} also reported results on Q-learning behavior in a Cournot oligopoly game where they found a slight tendency towards collusive behavior in their simulation study. \citeauthor{xu2020reinforcement} incorporated memory and imitation with RL. Firms do not have any information about market demand, but they can observe quantities produced by the other firms and their profits. Unlike \cite{waltman2008q}, they use a continuous action space and parametric function approximation. They use three different environment settings, the first setting (Treatment 1 in their paper) resembles our models. They have used the same experimental setup as in \cite{waltman2008q}. Unlike our results, they observe convergence to Nash equilibrium for these settings. Both of these papers evaluate scalability up to a limited range, while our paper incorporates a more thorough investigation of scalability in terms of the number of firms and the size of action space. \cite{huck2004through} studied another trial-and-error method. In their work, firms do not have information about their rivals as well as the payoff function of the game. However, their method allows agents only to decrease or increase the production level. Therefore, there is a great deal of responsibility on firms to choose the starting quantity. They showed that full collusion can be achieved with their approach; however, they did not study the scalability of their approach.

There is a wide variety of work that uses RL for dynamic pricing problems. The market models considered in these works are diverse \cite{den2015dynamic}. \cite{kephart2000pseudo} use Q-learning in one of such settings, and \cite{kononen2006dynamic} use Q-learning with function approximation as well as policy gradient method. \cite{misra2019dynamic} proposed a multi-armed bandit based algorithm for a multi-period dynamic pricing problem, where firms face ambiguity. \cite{hansen2020algorithmic,trovo2015multi} studied the applicability of MAB algorithms on dynamic pricing problems and proposed variants of the UCB algorithm. \citeauthor{hansen2020algorithmic}~found that with static demand having low noise, it is possible to get collusive behavior in firms, but as the noise increases, the results become indistinguishable from Nash equilibrium. On contrary, we did not notice any significant changes in our outcomes due to increase in noise.

\section{Preliminaries}

\subsection{Cournot Oligopoly Model}
We consider a standard Cournot oligopoly model with $n$ firms which simultaneously produce identical/homogeneous products. Firm $i$ produces quantity $q_{i}\in Z$ with cost function $C_{i}$. Let $p$ denote linear demand function with an inverse demand equation as

\begin{equation}\label{eq1}
    p = max(v - w\sum_{i=1}^{n}q_{i},0)
\end{equation}

where $v>0$ and $w>0$ denote two parameters. Firms have constant marginal cost. Firm $i$'s total cost is, $C_{i} = cq_{i}$, where $c$ is the constant marginal cost. The profit of a firm $i$ is calculated as

\begin{equation}\label{eq2}
\pi_{i} = p q_{i} - C_{i} \ for \ i=1,\dots,n
\end{equation}

In addition to demand function given in eq. \ref{eq1}, we also consider Stochastic (Multiplicative) demand function \cite{karlin1962prices,petruzzi1999pricing} as follows (taken from \cite{huang2013demand})
\begin{equation}\label{eq3}
p = max(\beta(v - w\sum_{i=1}^{n}q_{i}),0)
\end{equation}
where $\beta$ is the noise sampled from normal distribution (negative values of $\beta$ are ignored).

\subsection{Equilibria in Cournot Oligopoly Models}
Following definitions are for models with symmetric firms.

\begin{definition}
The \textbf{Cournot (Nash) Equilibrium} is obtained if each firm chooses the production level that maximizes its profit, given the production levels of its competitors. No firm wishes to unilaterally change its output level when the other firms produce the output levels assigned to them in the equilibrium. Firms individually maximize their profit; they do not maximize their joint profit. The resulting joint production level is $\frac{(v-c)n}{w(n+1)}$

\end{definition}

\begin{definition}
\textbf{Walrasian Equilibrium} is obtained if firms are not aware of their influence on the market price, and therefore behave as price takers. They adopt Walrasian rule and produce Walrasian quantity. The Walrasian rule is based on the assumption that a firm acting as price taker decides next-period output maximizing its profit. The resulting quantity dynamics leads to a dynamic equation that allows the Walrasian equilibrium output as the unique steady state \cite{radi2017walrasian}. The resulting joint production level is $\frac{(v-c)}{w}$

\end{definition}

\begin{definition}
In \textbf{Collusive Equilibrium}, firms form a cartel and maximize the joint profit. They produce a smaller quantity than the quantity that maximizes their individual profit. Hence, they have an incentive to increase their production levels. The resulting joint production level is $\frac{(v-c)}{2w}$

\end{definition}

\subsection{Multi-agent Multi-armed Bandit Framework}

We consider the firms in repeated Cournot game as the agents operating in a MA-MAB framework with their arms representing the different production choices. Agents do not have any knowledge of the environment; agents only have knowledge about their own payoffs/rewards after taking an action. The action space is discrete and ordered; every action represents a production level choice, and the action set consists of all the integer values in a specific range.

Let $n$ be the number of agents in our MA-MAB setting. As agents are independent and autonomous, they each face a separate stochastic MAB with $L$ arms. Each agent's goal is to maximize their individual reward over a long (possibly infinite) time horizon $T$. For a game played at time step $t$, the reward is same as the profit, which is calculated using eq. \ref{eq2}. From eq. \ref{eq2}, we can see that reward depends on individual agent's production as well as total industry output at time step $t$. Since all the agents are learning simultaneously, these bandits are not perfectly stochastic MAB, and the means and variances associated with arms are unknown to the agents.

\section{Methods}

We have modeled the repeated Cournot oligopoly game as a MA-MAB setting. Each firm in the market faces separate stochastic MAB. We have also explained that these are not perfect stochastic MABs, and the means and variances associated with arms are unknown. We feel that this is important to investigate since, as stated in~\cite{lattimore2020bandit}, perfectly stochastic MABs cannot necessarily be expected in the real world.

We initially analyzed the applicability of three well-known multi-armed bandit algorithms: $\epsilon$-greedy, UCB (Upper Confidence Bound), and Thompson sampling. Most of the literature deals with Bernoulli bandits; however, we are dealing with normal bandits with unknown means and variances associated with the arms. We found that the UCB-Normal algorithm \cite{cowan2017normal} which is designed for such scenarios, fails to converge. Thompson sampling proposed in \cite{honda2014optimality} is also designed to handle such scenarios; however, the outcomes depend on priors. As it is required to set appropriate priors, we cannot use it for our models. We think that $\epsilon$-greedy is the only one among these three algorithms which can work with our assumptions of the lack of knowledge. We propose two novel approaches specifically to deal with the ordered action space in the Cournot games. Both of these approaches use $\epsilon$-greedy as a sub-routine.

\subsection{$\epsilon$-Greedy}
$\epsilon$-Greedy \cite{sutton1998reinforcement} is a common method for balancing exploration and exploitation trade-offs. At each time step $t=1,2,..$, an agent chooses random action with probability $\epsilon$ or otherwise chooses the action with the highest empirical mean. The empirical mean of an action $a$ is often referred to as that action's $Q$-value and is denoted, $Q(a)$. In our formulation, $Q(a)$ is the running average of rewards obtained by choosing an action.

A linear bound on the expected regret can be achieved with constant $\epsilon$. For variant of the algorithm where $\epsilon$ decreases with time, \cite{cesa1998finite} proved poly-logarithmic bounds. However, \cite{vermorel2005multi} did not find any practical advantage to using these methods in their empirical study. Here, we are referring to regret as it is used conventionally in single agent stochastic MAB problems \cite{kuleshov2014algorithms,lattimore2020bandit}.

\subsection{$\epsilon$-Greedy with Hierarchical Learning ($\epsilon$-greedy+HL)} 

We propose a new approach similar to hierarchical learning. This approach makes use of the fact that the action set is ordered in Cournot games. Learning happens in multiple phases. In each phase, the $\epsilon$-Greedy algorithm is applied to certain arms. In the initial phase, the whole action set is divided into $K$ equally-sized ranges (buckets). Each action range/bucket is then considered as a single arm. When such an arm is pulled, the action is chosen uniformly at random from the bucket associated with that arm. Each phase ends when the same arm/bucket is chosen during exploitation mode for some pre-specified number of time steps. At the end of each phase, the arm with the highest $Q$-value is selected, and the associated bucket is further split into smaller-sized buckets. These new smaller buckets are then considered as arms for the next phase of the learning. This continues until the bucket cannot be further divided. In this way, under-performing arms are eliminated, and the focus of learning shifts to high-performing arms. Firms can select the value of $K$ as per their choice. This learning process can also be visualized as the search for the best action in $K$-ary tree.

\subsection{$\epsilon$-Greedy with Elimination of Arms ($\epsilon$-greedy+EL)}

With hierarchical learning, firms need to make partitions of their action space, which is a crucial but delicate task. To do it efficiently, firms may need some information about the environment. Wrong partitioning may lead to losing out on optimal actions, i.e., a seemingly optimal bucket may contain sub-optimal actions. For the circumstances where firms may not want to take the responsibility of partitioning the action space, we propose another novel approach that is based on eliminating, rather than partitioning, the actions.

This approach also relies on $\epsilon$-greedy as an underlying mechanism, and similar to hierarchical learning, works in multiple phases. Unlike previous approach, the initial phase starts with considering each action as an arm. Each phase ends when the same action $a$ is chosen during exploitation mode for some pre-specified amount of time steps. At the end of each phase, an action $a$ with maximum $Q$-value is selected. Let $m$ be the size of the action space. If available, $m/4$ number of actions on either side of action $a$ on the ordered scale are kept in the action space, and other actions are eliminated. This continues until $m <= 3$.

\section{Experiments and Results}

In experiments, we empirically evaluate all three methods discussed above, i.e., $\epsilon$-greedy, $\epsilon$-greedy+HL, and $\epsilon$-greedy+EL. We utilize simulations similar to the ones used by \cite{waltman2008q}. We further explore the scalability of our methods by running simulations with models consisting of up to $100$ firms and up to $500$ actions. We run every simulation $100$ times; here, simulation means the entire training process that lasts for several time steps. For each simulation, we compute the final outcome as the average value over last $100$ time steps. We consider both deterministic and stochastic versions of inverse demand function, and also, models with both symmetric and asymmetric firms.

\subsubsection{Evaluation Measures}

We aim to investigate the resulting equilibrium that Cournot models achieve after applying the proposed approaches. We compare joint quantities and joint profits obtained with baseline $\epsilon$-greedy as well as our proposed approaches. Moreover, we do an empirical analysis of the regrets. The environment evolves over time as agents learn. As a result, the optimal action to take changes from an individual agent's perspective. Also, there is no way to foretell the convergence of other agents' strategies to any specific action or policy. Therefore, we cannot calculate regret for individual agents the way it is calculated for typical stochastic MAB problems. However, we consider that the firms always strive for the maximum profit; this can only be achieved if firms form a cartel, i.e., full collusion. Therefore, despite agents being independent and autonomous, we calculate joint cumulative regret for the whole system. It is calculated by subtracting the joint cumulative profit that is actually obtained from the profit firms could have obtained by acting as a cartel. In graphs, the lines are plotted using mean values, and the fillers encompassing those lines are plotted using $25th$ and $75th$ percentiles.

\subsubsection{Parameters}
The exploration rate, $\epsilon$, is one of the main hyper-parameters in $\epsilon$-greedy approaches. In the reported experiments, we use the exploration rate $\epsilon=0.1$ for all three approaches. We also tried other exploration rates, and found that the final outcomes are robust to different exploration rates; however, an increase in joint cumulative regret can be seen with higher exploration rates. Moreover, decay in the exploration rate does not make any difference in outcomes or the convergence rate. We use the running average of rewards as the Q-values; therefore, the learning rate is not needed. Instead of learning for a pre-specified amount of steps, agents learn until they consistently pick a particular action in exploitation mode for the specific number of steps. For $\epsilon$-greedy+HL and $\epsilon$-greedy+EL, agents have to pick the same arm in exploitation mode for $100$ consecutive steps in each phase of the method. However, for baseline $\epsilon$-greedy, agents have to pick the same arm in exploitation mode for $1000$ consecutive steps. This difference is because there is no reduction in the size of the action space for baseline $\epsilon$-greedy, and it takes more steps to get to the steady choice in multiple simulation runs. These choices apply to all types of simulations discussed below. For $\epsilon$-greedy+HL, we use $K$ (number of partitions) to be $3$. The outcomes in terms of equilibrium are robust to the choice of $K$; nonetheless, it can affect joint cumulative regret as the number of phases in the algorithm and total time steps depend on this parameter.

\subsection{Small-scale Simulations}
\begin{figure*}
     \centering
    \begin{subfigure}[b]{0.33\textwidth}
    \centering
    \includegraphics[width=\textwidth]{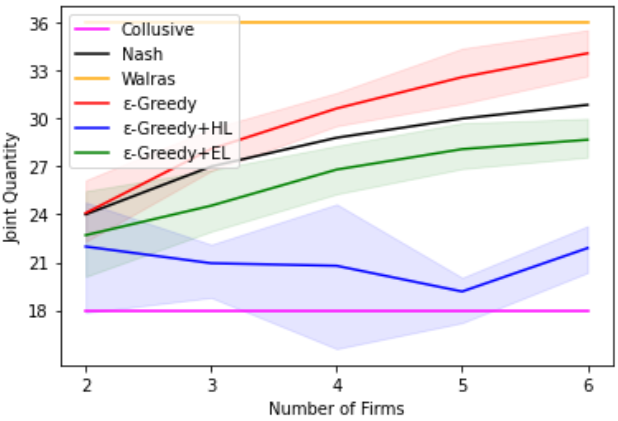}
    \caption{Comparison of \\ joint quantities}
    \label{fig1a}
    \end{subfigure}
    \hfill
    \begin{subfigure}[b]{0.33\textwidth}
    \centering
    \includegraphics[width=\textwidth]{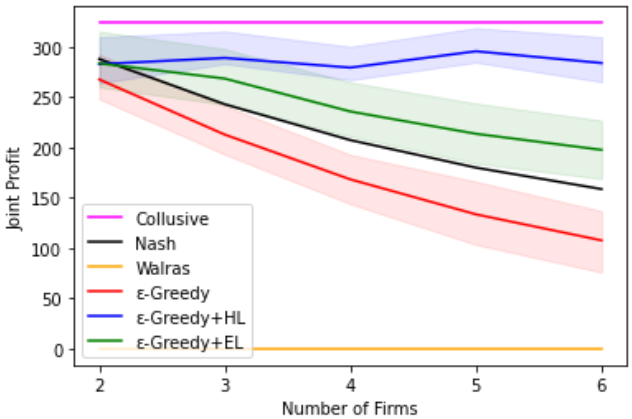}
    \caption{Comparison of \\ joint profits}
    \label{fig1b}
    \end{subfigure}
    \hfill
    \begin{subfigure}[b]{0.33\textwidth}
    \centering
    \includegraphics[width=\textwidth]{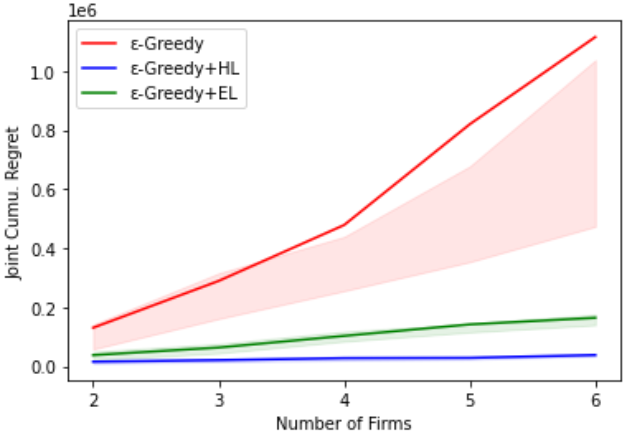}
    \caption{Comparison of \\ joint cumu. regrets}
    \label{fig1c}
    \end{subfigure}
\caption{For small-scale simulations, comparisons of different measures obtained by $\epsilon$-greedy, $\epsilon$-greedy+HL, and $\epsilon$-greedy+EL algorithm, along with collusive, Nash and Walrasian equilibrium. Cournot model has symmetric firms with $v=40$, $w=1$, and $c=4$.}
\label{}
\end{figure*}

We first evaluate our approaches using the same simulations as those used by \cite{waltman2008q}. They used Cournot models with symmetric firms in their simulations. They used $v=40$, $w=1$ as constants in an inverse demand function (eq. \ref{eq1}), along with constant marginal cost, $c=4$. The action space is discrete, where agents can choose a production quantity between $0-40$. They ran simulations with the total number of firms varying from $2$ to $6$. The results of these simulations are shown in fig \ref{fig1a}, \ref{fig1b}, and \ref{fig1c} for joint quantities, joint profits, and joint cumulative regrets, respectively. From fig \ref{fig1a} and \ref{fig1b}, we can see that with $\epsilon$-greedy, the outcomes mostly converge somewhere between Nash and Walras equilibrium; however, for model with $2$ firms, output coincides with Nash equilibrium in fig \ref{fig1a}. For both $\epsilon$-greedy+HL and $\epsilon$-greedy+EL, outcomes are collusive. However, with $\epsilon$-greedy+HL those are closer to collusive equilibrium. We consider an outcome to be \textit{collusive} when it is anywhere in between Nash and collusive equilibrium; therefore the degree of collusion may vary based on its closeness to collusive equilibrium. The outcomes seen in the graphs of joint quantities may not be perfectly reflected in the corresponding graphs of joint profits. This is a consistent outcome that can be seen throughout our experiments. We think that this is because we allow profits/rewards to be negative in our models. Firms lose the production costs regardless of the market price. Because of this, there is more variance in joint profits obtained through different simulations. As we take the average results of $100$ different simulations, it is reflected in the graphs.

From fig \ref{fig1c}, we can see that the $\epsilon$-greedy+HL method obtains lowest regret, followed by $\epsilon$-greedy+EL, and $\epsilon$-greedy obtains very high regret in comparison to the other two methods. One of the reasons is that our proposed approaches take fewer steps to converge than $\epsilon$-greedy. $\epsilon$-greedy+HL on average takes only $300$ steps while $\epsilon$-greedy+EL takes around $650$ steps. Graph comparing total time steps taken by these three approaches (shown in \ref{fig4a}) is very similar to joint cumulative regrets' graph shown in fig \ref{fig1c}. In addition to this, we think that the more collusive outcomes and rapid decrease in action space size causes $\epsilon$-greedy+HL to have lower regret than $\epsilon$-greedy+EL.

\begin{figure*}
    \centering
    \begin{subfigure}[b]{0.33\textwidth}
    \centering
    \includegraphics[width=\textwidth]{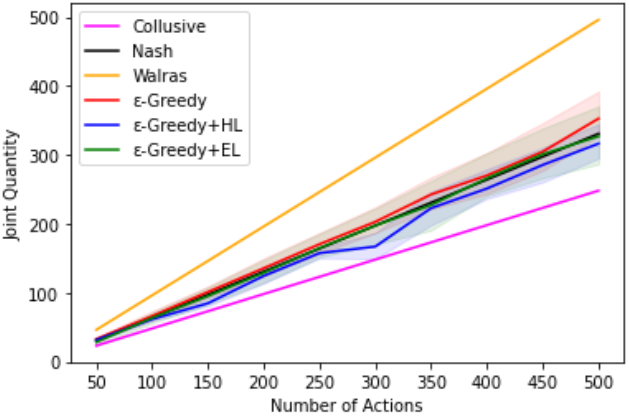}
    \caption{Comparison of joint quantities}
    \label{fig2a}
    \end{subfigure}
    \hfill
    \begin{subfigure}[b]{0.33\textwidth}
    \centering
    \includegraphics[width=\textwidth]{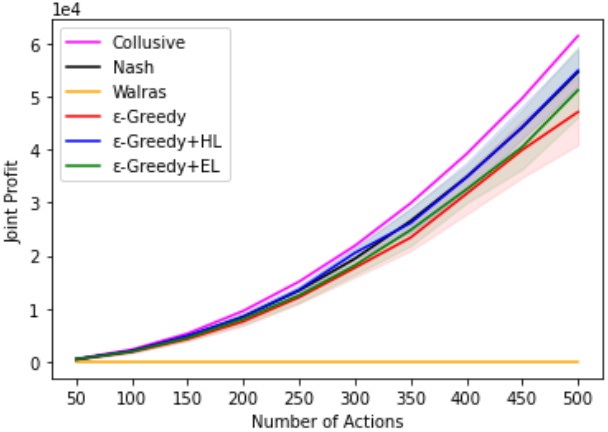}
    \caption{Comparison of joint profits}
    \label{fig2b}
    \end{subfigure}
    \hfill
    \begin{subfigure}[b]{0.33\textwidth}
    \centering
    \includegraphics[width=\textwidth]{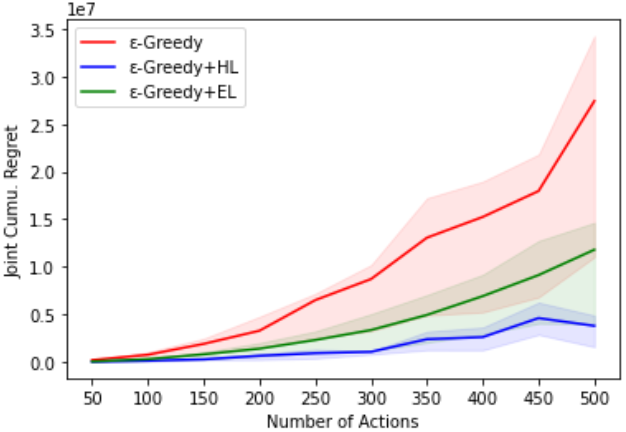}
    \caption{Comparison of joint cumu. regrets}
    \label{fig2c}
    \end{subfigure}

    \caption{For simulations with scaled actions, comparisons of different measures obtained by $\epsilon$-greedy, $\epsilon$-greedy+HL, and $\epsilon$-greedy+EL algorithm, along with collusive, Nash and Walrasian equilibrium. Cournot model has symmetric firms.The size of action space, $S$, varies from $50$ to $500$, with $v=S$ and $w=1$, and $c=4$. }
    \label{}
\end{figure*}

\begin{figure*}
    \centering
    \begin{subfigure}[b]{0.33\textwidth}
    \centering
    \includegraphics[width=\textwidth]{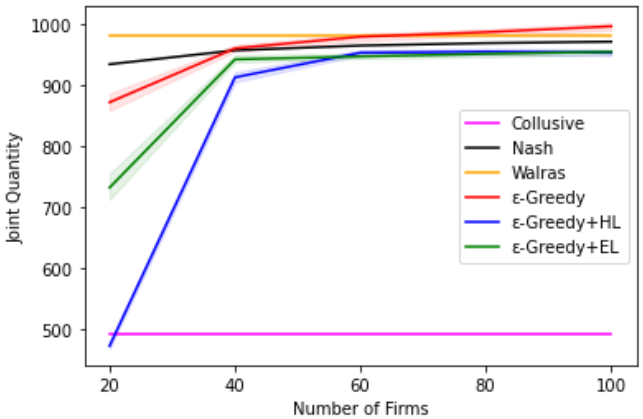}
    \caption{Comparison of joint quantities}
    \label{fig3a}
    \end{subfigure}
    \hfill
    \begin{subfigure}[b]{0.33\textwidth}
    \centering
    \includegraphics[width=\textwidth]{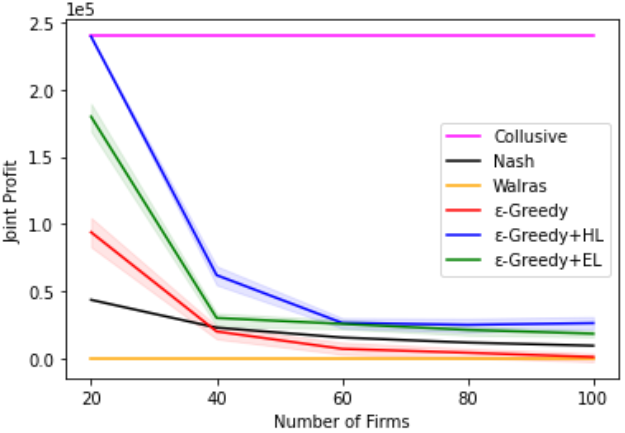}
    \caption{Comparison of joint profits}
    \label{fig3b}
    \end{subfigure}
    \hfill
    \begin{subfigure}[b]{0.33\textwidth}
    \centering
    \includegraphics[width=\textwidth]{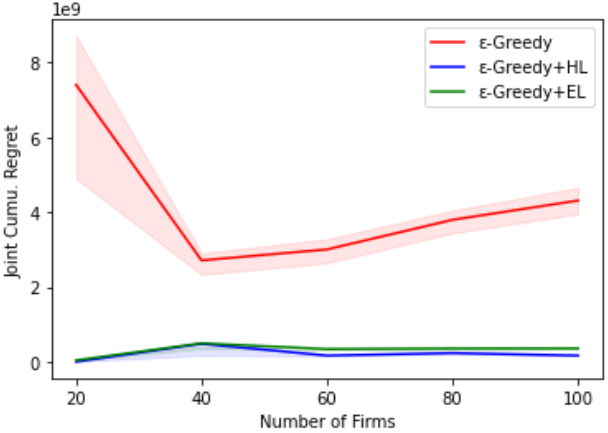}
    \caption{Comparison of joint cumu. regrets}
    \label{fig3c}
    \end{subfigure}
    \caption{For simulations with scaled number of agents, comparisons of different measures obtained by $\epsilon$-greedy, $\epsilon$-greedy+HL, and $\epsilon$-greedy+EL algorithm, along with collusive, Nash and Walrasian equilibrium. Cournot model has symmetric firms with $v=1000$ and $w=1$, and $c=20$.}
    \label{}
\end{figure*}

\begin{figure*}
    \centering

    \begin{subfigure}[b]{0.33\textwidth}
    \centering
    \includegraphics[width=\textwidth]{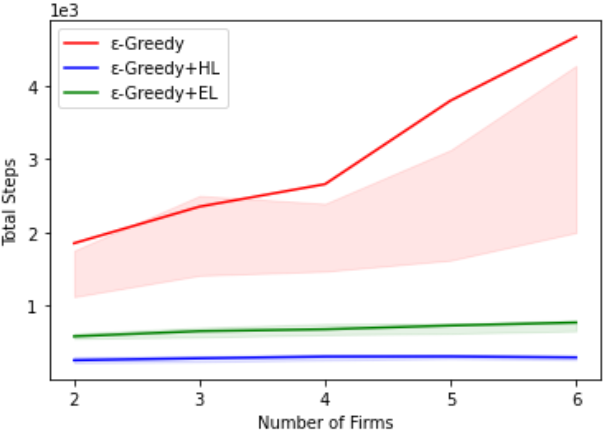}
    \caption{Small-scale \\ simulations}
    \label{fig4a}
    \end{subfigure}
    \hfill
    \begin{subfigure}[b]{0.33\textwidth}
    \centering
    \includegraphics[width=\textwidth]{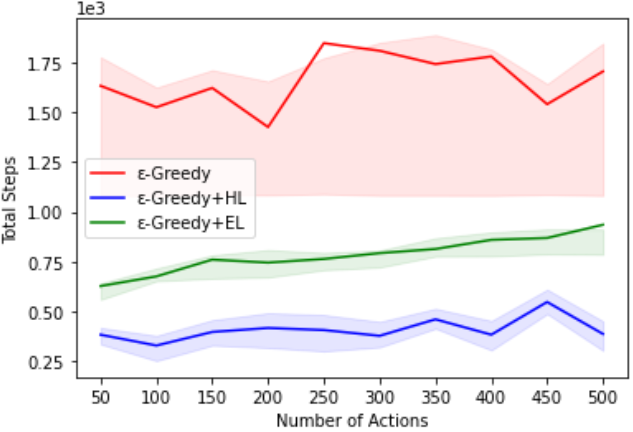}
    \caption{Large-scale simulations \\ (scaled actions)}
    \label{fig4b}
    \end{subfigure}
    \hfill
    \begin{subfigure}[b]{0.33\textwidth}
    \centering
    \includegraphics[width=\textwidth]{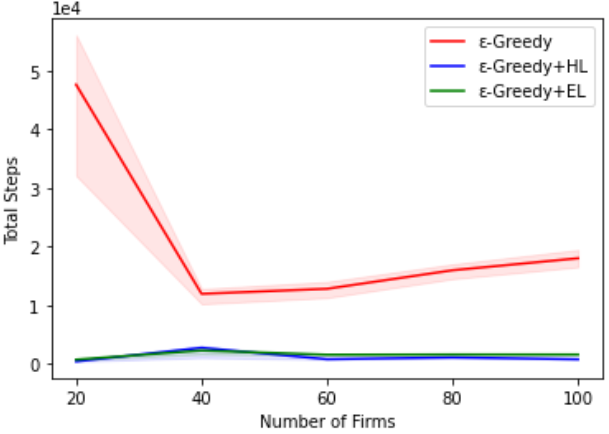}
    \caption{Large-scale simulations \\ (scaled agents)}
    \label{fig4c}
    \end{subfigure}

\caption{Comparison of average total time steps taken by agents to converge}
\label{}
\end{figure*}

\subsection{Large-scale Simulations}

\subsubsection{Scaling action space:}

Firms with high production capacities can exist in the market. With high production capacity, more actions, i.e., production level choices, are available to the firms. Here, we use the Cournot duopoly model consisting of $2$ symmetric firms. However, the size of action space, $S$, varies from $50$ to $500$. Available actions in the specific model are in the range $0-S$. For the meaningful exploration of action space, the market demand also varies according to the size of the action space, such that the value of parameter $v$ in the demand function is equal to the size of action space, $S$. From fig \ref{fig2a}, we can see that most of the outcomes converge to Nash equilibrium. However, for $\epsilon$-greedy+HL outcomes are somewhat collusive, and for $\epsilon$-greedy, those are bit off from Nash equilibrium and lies between Nash and Walras equilibrium.

For the reasons discussed in previous section, joint profits shown in fig \ref{fig2b} are slightly different from their counterpart joint quantities in fig \ref{fig2a}. As shown in fig \ref{fig2c}, the regret with $\epsilon$-greedy is larger than that for $\epsilon$-greedy+EL, which is, in turn, larger than the regret for $\epsilon$-greedy+HL. In fig \ref{fig2c}, joint cumulative regret increases with the size of action space for all three approaches, but this pattern is not seen with the total steps required to converge (from \ref{fig4b}); for all approaches, total steps required to converge do not differ much with varying action space size; nonetheless, they are directly proportional to joint regrets.

\subsubsection{Scaling number of firms:}

Cournot models with a large number of firms have been rarely explored in the literature. We simulate models consisting of $20$, $40$, $60$, $80$, and $100$ total number of firms. Here, we consider the demand function with $v=1000$ and $w=1$, along with $c=20$. The action choices range from $0$ to $50$. The results for joint quantity, joint profit, and joint cumulative regret are shown in figures \ref{fig3a}, \ref{fig3b}, and \ref{fig3c}, respectively. With increasing number of firms, Nash equilibrium approaches Walrasian equilibrium. In fig \ref{fig3a}, for $\epsilon$-greedy we can see that the outcome is collusive for model with $20$ firms; with $40$ firms, it coincides with Nash equilibrium; for models with $60$ and $80$ firms, it seems to coincide with Walras equilibrium and interestingly, outcomes seems to be slightly more than Walras equilibrium for model with $100$ firms. Therefore, we can say that $\epsilon$-greedy performs very poorly for models with large number of firms and begins to break down when the total number of firms is greater than $80$. Surprisingly, if we compare its performance in terms of joint profits, it is better in the sense that it does not fall below Walras equilibrium for model with $100$ firms. This is unlike the results for previous sections where joint profits tend to be worse than their counterparts in joint quantities (if we assume that collusive $\succ$ Nash $\succ$ Walras).

$\epsilon$-greedy+HL and $\epsilon$-greedy+EL both obtain collusive results in these simulations. For models with comparatively smaller number of firms ($20$ and $40$), $\epsilon$-greedy+HL seems to be more collusive; however, outcomes of both approaches are similar for larger models. Unlike previous results, $\epsilon$-greedy+HL and $\epsilon$-greedy+EL methods obtain nearly the same regrets. Not surprisingly, we found that both $\epsilon$-greedy+HL and $\epsilon$-greedy+EL run for almost equal time steps (shown in \ref{fig4c}). Overall, the graph for total time steps is very similar to fig \ref{fig3c} showing joint cumulative regrets. Interestingly, even though the model with $20$ firms converge to collusive outcomes, the regret obtained by $\epsilon$-greedy in this case is higher than other larger models. As the demand and size of action space stay the same across all models with varying number of firms, agents in smaller models have more viable action choices that give positive payoffs/rewards. It causes a delay in converging to the final outcome and hence more exploration, which in turn causes more regret. From both types of large-scale simulations, we can see that having more viable action choices leads to more regret.

\subsubsection{Stochastic demand function}

As mentioned in the preliminaries, the stochastic demand function (eq. \ref{eq3}) is obtained by multiplying the inverse demand function (eq. \ref{eq1}) with noise. This noise is derived from a normal distribution. We evaluate two types of models with stochastic demand. In the first one, noise is derived from a normal distribution having mean, $\mu=1$, and standard deviation, $\sigma=0.1$. For the second type of model, we increase $\sigma$ to $0.2$. We found no significant difference between the outcomes for both the cases, and the results are very similar to the ones obtained by having deterministic demand. We examined the results for both small-scale and large-scale simulations. For both types of simulations, there is no difference between outcomes in terms of joint quantity and joint profit; this is true for all three approaches. For small-scale simulations, with $\epsilon$-greedy, the outcomes obtained for models with stochastic demand are worse than those obtained with deterministic demand in terms of joint cumulative regrets; in most of the cases, more stochasticity leads to worse outcome. However, this difference lessened in the results obtained by $\epsilon$-greedy+HL and $\epsilon$-greedy+EL, which are almost same for all types of models. It shows that our proposed approaches can handle stochasticity in the demand function. Surprisingly, for large-scale simulations, in some cases, joint cumulative regret obtained for models with static demand is slightly more than that obtained for models with stochastic demand. Here, stochasticity might have helped to find better actions.

\subsubsection{Asymmetric firms}
We also study Cournot models with asymmetric firms, i.e., firms with different marginal costs. In simulations, we use Cournot duopoly models with two firms having different marginal costs. These costs vary in different simulations. We used the same model parameters as those used in small-scale simulations. We have $v=40$ and $w=1$ as constants in an inverse demand function (eq. \ref{eq1}), but the constant marginal cost $c$ is different for two firms. Let $c1$ and $c2$ be the marginal costs for two firms, then $c1$ lies in range $1-3$ and $c2$ in range $3-5$. Overall, the results (shown in \ref{tab:table1}) are similar to those with symmetric firms. $\epsilon$-greedy mostly converges to outcomes worse than Nash equilibrium. However, for $\epsilon$-greedy+HL and $\epsilon$-greedy+EL, most of the outcomes are collusive; more collusive with $\epsilon$-greedy+HL than with $\epsilon$-greedy+EL.

\begin{table*}

\centering
\setlength\extrarowheight{10pt}
\begin{tabular}{|c|c|c|c|c|c|}
\hline

\multicolumn{2}{|c|}{$(C1,C2)$} &
\shortstack{Nash\\Equilibrium} &
\shortstack{$\epsilon$-greedy} &
\shortstack{$\epsilon$-greedy \\ + HL} &
\shortstack{$\epsilon$-greedy \\ + EL} 
 \\
\hline

\multirow{2}{*}{(1, 2)}	&	Q	&	[13.3, 12.3]	&	[13.8±4.3, 12.4±3.9]	&	[11.3±2.9, 10.8±2.7]	&	[11.4±3.9, 12.7±4.7]	\\
\cline{2-6}
	&	P	&	[177.8, 152.2]	&	[170.6±49.0, 143.2±49.3]	&	[181.6±35.0, 164.2±27.7]	&	[163.9±38.2, 164.5±30.3]	\\
\hline
\multirow{2}{*}{(1, 3)}	&	Q	&	[13.7, 11.7]	&	[13.7±5.2, 12.3±5.5]	&	[11.8±3.7, 10.0±3.0]	&	[12.8±3.1, 10.5±4.1]	\\
\cline{2-6}
	&	P	&	[186.7, 136.1]	&	[166.8±69.5, 122.3±66.5]	&	[196.9±45.3, 142.4±43.0]	&	[189.2±29.4, 138.3±36.0]	\\
\hline
\multirow{2}{*}{(1, 4)}	&	Q	&	[14.0, 11.0]	&	[13.5±4.6, 12.1±5.2]	&	[11.9±3.2, 9.7±2.4]	&	[12.8±3.6, 10.1±3.2]	\\
\cline{2-6}
	&	P	&	[196.0, 121.0]	&	[177.2±62.1, 117.6±58.8]	&	[199.6±39.1, 129.6±43.5]	&	[196.7±23.1, 125.3±31.8]	\\
\hline
\multirow{2}{*}{(1, 5)}	&	Q	&	[14.3, 10.3]	&	[13.5±4.6, 11.3±4.8]	&	[12.6±2.4, 8.4±3.1]	&	[12.4±2.7, 10.0±2.1]	\\
\cline{2-6}
	&	P	&	[205.5, 106.8]	&	[180.9±59.6, 107.1±56.4]	&	[216.0±47.5, 109.5±34.4]	&	[194.6±24.5, 122.9±31.9]	\\
\hline
\multirow{2}{*}{(2, 2)}	&	Q	&	[12.7, 12.7]	&	[12.5±5.1, 12.9±4.8]	&	[10.9±2.6, 10.9±2.8]	&	[11.7±3.4, 11.8±3.8]	\\
\cline{2-6}
	&	P	&	[160.4, 160.4]	&	[149.4±63.6, 154.0±60.1]	&	[170.2±44.6, 167.8±35.1]	&	[162.7±30.2, 160.7±27.7]	\\
\hline
\multirow{2}{*}{(2, 3)}	&	Q	&	[13.0, 12.0]	&	[13.1±5.1, 13.2±5.7]	&	[11.6±3.0, 9.6±2.2]	&	[12.8±2.7, 11.1±3.3]	\\
\cline{2-6}
	&	P	&	[169.0, 144.0]	&	[148.6±67.1, 133.2±67.5]	&	[186.9±43.8, 142.8±36.2]	&	[173.9±26.4, 141.3±35.0]	\\
\hline
\multirow{2}{*}{(2, 4)}	&	Q	&	[13.3, 11.3]	&	[13.6±5.0, 11.2±4.9]	&	[11.9±3.3, 10.0±3.6]	&	[12.3±3.4, 10.7±2.1]	\\
\cline{2-6}
	&	P	&	[177.8, 128.5]	&	[167.5±67.2, 115.8±60.1]	&	[185.9±44.3, 131.8±31.7]	&	[175.9±36.0, 133.6±33.0]	\\
\hline
\multirow{2}{*}{(2, 5)}	&	Q	&	[13.7, 10.7]	&	[13.3±4.7, 12.1±4.7]	&	[12.7±3.9, 7.9±3.2]	&	[12.6±2.6, 9.6±2.7]	\\
\cline{2-6}
	&	P	&	[186.7, 113.7]	&	[160.1±62.2, 106.5±64.1]	&	[211.8±49.8, 103.0±32.4]	&	[186.1±30.4, 118.2±27.0]	\\
\hline
\multirow{2}{*}{(3, 2)}	&	Q	&	[12.0, 13.0]	&	[12.9±6.1, 13.0±4.7]	&	[9.8±3.4, 12.2±2.6]	&	[11.6±2.6, 11.9±2.3]	\\
\cline{2-6}
	&	P	&	[144.0, 169.0]	&	[130.5±66.7, 153.7±66.3]	&	[141.8±40.2, 186.9±36.0]	&	[147.4±30.3, 164.9±25.5]	\\
\hline
\multirow{2}{*}{(3, 3)}	&	Q	&	[12.3, 12.3]	&	[13.5±6.1, 12.3±4.4]	&	[9.7±3.5, 11.3±2.4]	&	[11.5±2.8, 11.6±2.1]	\\
\cline{2-6}
	&	P	&	[152.2, 152.2]	&	[135.9±68.0, 135.4±61.6]	&	[145.5±29.9, 172.0±31.9]	&	[153.1±29.3, 154.1±28.4]	\\
\hline
\multirow{2}{*}{(3, 4)}	&	Q	&	[12.7, 11.7]	&	[12.9±4.5, 11.9±5.1]	&	[11.6±2.3, 8.9±2.1]	&	[11.4±2.4, 11.4±2.3]	\\
\cline{2-6}
	&	P	&	[160.4, 136.1]	&	[149.6±62.4, 121.3±64.0]	&	[179.9±29.2, 130.1±27.2]	&	[152.6±23.9, 142.8±31.9]	\\
\hline
\multirow{2}{*}{(3, 5)}	&	Q	&	[13.0, 11.0]	&	[13.3±5.4, 11.3±3.5]	&	[11.5±2.9, 9.0±3.0]	&	[12.2±2.2, 10.2±2.8]	\\
\cline{2-6}
	&	P	&	[169.0, 121.0]	&	[149.8±58.4, 115.6±54.7]	&	[180.8±23.0, 121.1±21.3]	&	[168.5±27.9, 120.5±23.6]	\\

\hline
\end{tabular}
\caption{Results of computer simulations for asymmetric Cournot duopoly model with firms having different marginal costs $C1$ and $C2$; Demand Function: $v-wU$; $U$=Total industry output; $v=1000$, $w=1$. Q=Quantity; P=Profit. Results are in format $[q_{1},q_{2}]$ for quantities and $[p_{1},p_{2}]$ for profits; $q_{i}$ and $p_{i}$ represents quantity and profit for firm $i$, respectively.}

\label{tab:table1}
\end{table*}

\section{Conclusion and Future Work}

We have investigated the modeling of Cournot games in MA-MAB setting, especially when the firms have no knowledge of the market. Using MA-MAB setting allows for the implementation of approaches based on exploration-exploration paradigm. To deal with ordered action sets, we proposed two novel approaches, which are extensions of $\epsilon$-greedy algorithm. Given the assumption of static demand, proposed methods optimize the exploration by reducing the action space. $\epsilon$-greedy+HL performs best in terms of joint cumulative regret but comes with the additional responsibility of deciding the number of partitions in action space. With $\epsilon$-greedy+EL, there is no such burden, but it may cause more regret than $\epsilon$-greedy+HL. Our proposed approaches converge much faster than baseline $\epsilon$-greedy method. $\epsilon$-greedy rarely produces collusive outcomes, but mostly obtains Nash quantity or any state between Nash and Walrasian outcome. Our proposed approaches mostly obtain somewhat collusive outcomes for all kinds of simulations, although full collusion usually does not emerge. We think that these results are supportive of the concern over online algorithms being collusive without any external intervention \cite{timo2021ethics}. In future, we would like to work with Cournot models having non-stationary demand functions, and random entry or exit allowed to the firms.

\bibliographystyle{unsrtnat}
\bibliography{references}
\end{document}